\documentclass[]{article}
\usepackage[T1]{fontenc}
\usepackage{lmodern}
\usepackage{amssymb,amsmath}
\usepackage{ifxetex,ifluatex}
\usepackage{fixltx2e} % provides \textsubscript
\IfFileExists{upquote.sty}{\usepackage{upquote}}{}
\ifnum 0\ifxetex 1\fi\ifluatex 1\fi=0 % if pdftex
  \usepackage[utf8]{inputenc}
\else % if luatex or xelatex
  \ifxetex
    \usepackage{mathspec}
    \usepackage{xltxtra,xunicode}
  \else
    \usepackage{fontspec}
  \fi
  \defaultfontfeatures{Mapping=tex-text,Scale=MatchLowercase}
  
\fi
\IfFileExists{microtype.sty}{\usepackage{microtype}}{}
\usepackage{longtable,booktabs}
\usepackage{graphicx}
\makeatletter
\def\ScaleIfNeeded{%
  \ifdim\Gin@nat@width>\linewidth
    \linewidth
  \else
    \Gin@nat@width
  \fi
}
\makeatother
\let\Oldincludegraphics\includegraphics
{%
 \catcode`\@=11\relax%
 \gdef\includegraphics{\@ifnextchar[{\Oldincludegraphics}{\Oldincludegraphics[width=\ScaleIfNeeded]}}%
}%
\ifxetex
  \usepackage[setpagesize=false, % page size defined by xetex
              unicode=false, % unicode breaks when used with xetex
              xetex]{hyperref}
\else
  \usepackage[unicode=true]{hyperref}
\fi
\title{Public Spheres in Twitter- and Blogosphere. Evidence from the US.}
\author{Christoph Waldhauser, KDSS K Data Science Solutions}
\date{Paper presented at EPSA 2014}

\begin{document}
\maketitle

\section{Introduction}\label{introduction}

A major part of politics is the conveying of ideas. This is the function
of political communication. Of the multitude of ways this conveying can
take place, deliberation in a public sphere is one that Habermas (1994)
termed optimal. Both public spheres and deliberative democracy
presuppose an equality of all participants in a debate and the
superiority of good arguments. Presently, both features are not evident.

Large debates with many participants require a mediator to relay
messages from one participant to the next. To some extent, mass media
tries to fulfill that role. However, the use of mass media as mediator
comes at a cost: under the impression of the sheer volume of messages to
deliver, mass media needs to prioritize and select messages (Strömbäck
2008). This selection heuristic, however, is far from Habermas' ideal:
not all participants are equal and arguments not always rule (Meyer and
Hinchman 2002).

In contrast to mass media, social media---by definition---implements
equality of all users from the start and also places the selection of
messages under their control. It therefore has the potential to deliver
on both: fulfilling a role as a public sphere and promoting the
qualities of deliberative debate. In this contribution I establish a
baseline of social media's role as a public sphere, using recent Twitter
data.

This paper is organized as follows. I will first review the literature
pertaining to public spheres and their interaction with media, mass or
otherwise. I will then turn to describe the used data set and the
employed analytical methods. The next section will present the results
of my analysis. Thereafter I will discuss these results in light of the
theory reviewed above and close with some concluding remarks.

\section{Public Spheres \& Media}\label{public-spheres-media}

The original function of political communication was to facilitate the
finding of agreeable positions. Informed citizens were to exchange and
discuss ideas in order to discover these ideas' applicability to
reality. As societies quickly---quite literally---outgrew the
practicality of round tables and forum discussions, something Habermas
termed the public sphere (Habermas 1991; Habermas 2006) took
over.\footnote{Common features of public spheres identified in the
  literature are a mutual mode of communication and equally reciprocal
  acceptance of each other as legitimate participants (Koopmans and Erbe
  2004; Risse and Steeg 2003).}

The concept of a public sphere is embedded in the wider framework of
deliberative democracy. If the public sphere is the location where
political debates take place, deliberative democracy describes the mode
of these debates (Habermas 1994).

Cohen (2003) extends Habermas when identifying the common good as the
aim of all deliberation in deliberative democracy. His formal definition
entails five points: independence of association, self-referencing,
equality of ideas, rooted in reality, and reciprocal respect.

In his in-depth study of traces of deliberative democracy on the
Internet, Dahlberg (2001) uses a similar list but adds a sixth point to
it: reflexivity, that is the requirement to question one's own
standpoints.

From this list it becomes clear, that deliberative democracy not only
describes a practice, but a form of institution. Obviously, it is not a
role that a medium can fulfill. Even Dahlberg (2001) had to concede,
that the technology of that time simply was not up to providing for
institutionalized deliberative democracy on the Internet. However, it is
still possible to carve out the term of deliberative practices:
exchanging ideas and objective, reasonable, Dahlberg (2001) calls them
rational--critical, arguments in their support.

Thinking deliberative practices in concert with a public sphere makes
the sensible, fact-based argument its communicative hallmark feature. In
his introduction to a collection of essays on deliberative democracy,
Elster (1998) distills a common ground of contemporary definitions of
the deliberative practices part of deliberative democracy:

\begin{quote}
``decision making by means of arguments offered \emph{by} and \emph{to}
participants who are committed to the values of rationality and
impartiality'' (p.~8, emphasis in the original).
\end{quote}

It is this definition of deliberative practices that I will use here.
Deliberative practices serve to instill life into the hollowness of a
public sphere by describing the mode of communication that is conducive
to the functioning of a public sphere. This requires reasonable
arguments. As we shall see in the next section, they are to be found in
specific kinds of news stories. News stories, that are not always of the
prevailing kind.

\subsection{Media}\label{media}

In the modern public sphere, or spheres (Dahlgren 2005), it is still
citizens directly exchanging ideas and engaging in political debate.
However, Habermas' conception was an idealized one from the beginning.
Neither did citizens have the interest nor the talent or time to be
constantly engaged in debates. Rather, a mediator and a caste of
professional citizens emerged, the former distilling political messages
of the latter into a format that allowed even part-time citizens to---at
least on some level---partake in the debate. While this extends the
accessibility of the public sphere's debates to a larger crowd, there is
also another side to that bargain: the mediator might transform the
messages sent by other participants in a debate. And once there are more
messages fed into the mediator than can reasonably be processed and
relayed to others, the mediator will start selecting messages according
to its own rules. And this is, where trouble starts.

Schulz (2004) and Strömbäck (2008) summarize these troubles under the
term of mediatization. Here, mass media acts as the mediator and in that
role tampers with the political system in a number of ways. Foremost, it
selects messages to relay to others by its own media logic: newspaper
circulation figures and TV ratings, for example. Mass media uses these
numbers to gauge the profitability of any message. Will any given
message contribute to these figures and therefore increase the medium's
income, or not? Based on the answer to that question, a message will be
relayed, or not.

To further increase the value of a message, mass media will restyle the
message and potentially cripple its original meaning. This media logic
becomes internalized by politicians seeking to maximize their mass media
presence. This leads to the interesting phenomenon of shifting political
communication's attention from conveying ideas to reformulating ideas to
concur with this specific media logic.

The regime of mediatization is not necessarily detrimental to the
functioning of a democracy or even the bare exchange of ideas. As long
as the media logic captures the idea correctly and does not penalize the
complexity of ideas nor favor simplistic solutions, mediatization has
the potential of actually enriching public sphere debates. For example,
consider data journalism or fact-checking politicians' claims. Both
methods comply with a media logic that favors truth over simplicity,
therefore enticing politicians to communicate ideas truthfully and as
complex as necessary.

On the other hand, the vast majority of mass media follows a different
logic: here, shortness, entertainment and drama are elements that are of
greater importance than an idea's foundation in facts or civic vision
(Iyengar and Simon 2000; Iyengar, Luskin, and Fishkin 2004; Iyengar and
Kinder 2010; Strömbäck 2008). This media logic leads politicians to
offer populist solutions---quick and dirty, sellable to a broad audience
with short attention spans (Strömbäck and Esser 2009; Meyer and Hinchman
2002; Hjarvard 2008; Strömbäck 2008). This media logic cannot aptly
capture ideas anymore, and therefore leads to an erosion of the public
sphere.

Iyengar and Kinder (2010) argue that due to the advent of cable TV in
the 1990s, mass media is subjected to an heightened state of
competition. While up to the 1980s, only three networks supplied largely
identical news to a large audience, the exponential growth of media
outlets enabled citizens to opt-out of news. Or, if they still actively
listen to news, they can now choose which news they subscribe to. This
leads to media outlets producing news that is targeted specifically at
their (remaining) share of customers: highly partisan and appealing to
them.

In a political context, what exactly is appealing to consumers of news?
Iyengar, Norpoth, and Hahn (2004) answer this question clearly with
``horse race'`or episodic news, that is news that cover the current
state of the campaign, and not thematic content on the candidates'
positions. They understand episodicity and thematicity as opposing
concepts: the more episodic a piece of news is, the less thematic it can
be. While their contribution has numerous methodological flaws, they
convincingly put forth their argument of a decline of thematic content
due to competitive pressure among media outlets. This argument is also
in line with theory (Traugott and Lavrakas 2008), that would predict
these market effects to occur along with intrinsic journalistic
self-selection.

Considering the thematic--episodic content rift in terms of deliberative
practices allows for additional insights. If mass media function as a
mediator in Habermasian public spheres, then their debate contributions
should be based on reason and impartiality. Iyengar, Norpoth, and Hahn
(2004) give a textbook example for both kinds of contributions: a story
about a homeless woman and her plight; a story intended to serve as a
proxy for the fates of countless others. Touching as the story might be,
it is void of reason and does not serve to contribute to a rational
debate based on facts and impartial arguments. This episodic news story
is contrasted by a thematic one: thorough research of facts, causes and
consequences of homelessness; founded in reason and instructive in any
search for solutions, abstracting away individual stories for the sake
of the bigger picture.

As noted above, competition among media outlets has grown to tremendous
proportions. Since all competitors in the mass media game are commercial
enterprises and therefore immanently seeking to maximize their profit,
the dominant factor for selecting news is its market value. However,
this value does not refer to the amount a customer is willing to pay for
any particular piece of news. Rather, this figure correlates directly
with circulation figures or ratings as they govern the ad premiums a
media outlet can charge.

A factor that is not to be underestimated is the vicious circle this
mass media mediatization can lead to. Politicians vying for media
presence become then tempted to style their own messages in a media
compatible way. An example might be the Team Stronach's campaign for the
2013 general election in Austria; more specifically its reaction to a
high-profile international custody battle over a child. There, an
industrial magnate offered his private jet to the Austrian mother to
facilitate her legal struggle in Denmark. This episodic news story
worked very well and secured nationwide (tabloid) headlines for his
party and the generous offer. However, any political ideas on how
international custody disputes should be resolved in the future, any
abstract argument, was either not sent by Stronach's spokespersons or
not relayed by mass media. Therefore, the entire message did not
contribute to any deliberative practice.

In summary, there are several clues that point towards mass media
communication not always being true to its envisioned role in public
spheres and deliberative debates. The pressure from competition leads to
mass media selecting and transforming news to adhere to a style that
maximizes its market value (Blumler and Kavanagh 1999).

Social media is potentially different. There, users seek to maximize
their own value by providing content their respective audiences find
useful (Raymond 1999; Rheingold 1993; Pettersson and Karlström 2011).
Therefore, users would be expected to propagate content they themselves
believe their followers will find interesting. In an arena free from
traditional market pressures, this could lead to users selecting
thematic over episodic content.

\section{Data \& Method}\label{data-method}

In order to test the hypothesis put forth above, the last 3200 (the
maximum the Twitter API would allow) tweets emanating from the
Republican and Democratic party accounts, respectively, were harvested
on May 25th, 2014. As not both parties generate tweets at the same rate,
coverage is provided starting from 2013-02-13 for Democrats and
2013-07-31 for Republicans. Apparently, Democrats use their account much
more sparingly than Republicans. Given that @BarrackObama can also be
attributed to Democratic party control, this is of little surprise.

\begin{figure}[htbp]
\centering
\includegraphics{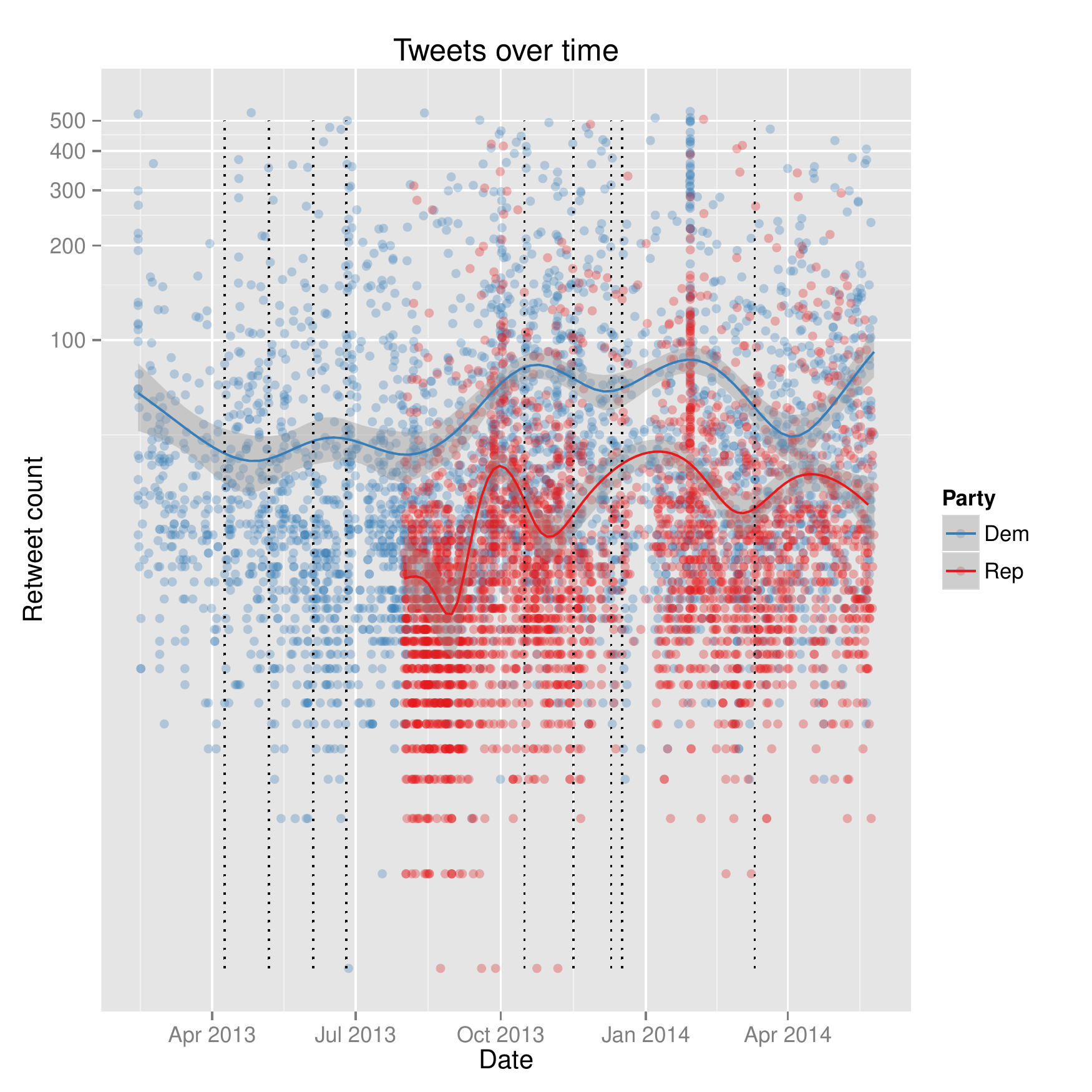}
\caption{\label{fig:overtime-plot}Development of retweets over time for
both parties with GAM-based trend line and elections to House or Senate
indicated by dotted verticals.}
\end{figure}

In general, Democrats enjoy more popularity on Twitter than Democrats.
This leads to Democrat's tweets being more often retweeted than their
Republican counterparts. On average, Democratic tweets were retweeted 60
times, while Republican ones only score half as many retweets on
average: 28. Since the range of retweets in enormous, with the most
popular tweet in the data set being retweeted 24,352 times\footnote{This
  tweet was a retweet itself of a message that Wendy Davis of Texas
  initially posted thanking Texans for support in stopping controversial
  Senate Bill 5 on abortions.}, the top 1 percent of retweeted tweets
were considered singular and therefore excluded from all
analysis\footnote{The top 1 percent of tweets range from 534 to above
  24,000 retweets. 64 tweets fell into that range and were excluded.}.
Figure \ref{fig:overtime-plot} describes the development of tweet
popularity over time.

\begin{figure}[htbp]
\centering
\includegraphics{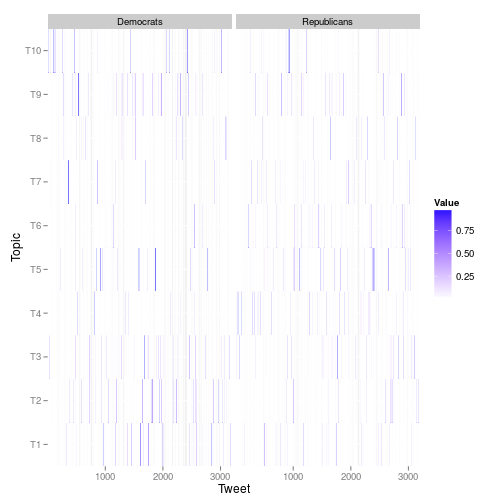}
\caption{\label{fig:topic-heatmap}Heatmap depicting the distribution of
topics across all tweets for both parties. The darker the spike, the
stronger the topic occurs in the tweet.}
\end{figure}

Of the 6,336 tweets that remained in the data set, 3,592 pointed to
resolvable websites. Those were harvested as well. The texts on those
websites were extracted, cleaned, normalized and fed into an LDA topic
model algorithm. The number of topics, 10, was empirically established.
Figure \ref{fig:topic-heatmap} gives an overview of topic distribution
among tweets.

The extracted topics were classified according to their aspect as either
episodic or thematic. A third ``aspect'' was necessitated by websites
that contained unparsable content, like pictures or content hidden by
AJAX techniques. These topics were then not considered any further.
Tables \ref{tab:dt} and \ref{tab:rt} give an overview over the extracted
topics and their rating as thematic or episodic.

\begin{longtable}[c]{@{}cll@{}}
\toprule\addlinespace
\begin{minipage}[b]{0.10\columnwidth}\centering
Topic
\end{minipage} & \begin{minipage}[b]{0.39\columnwidth}\raggedright
Content
\end{minipage} & \begin{minipage}[b]{0.15\columnwidth}\raggedright
Aspect
\end{minipage}
\\\addlinespace
\midrule\endhead
\begin{minipage}[t]{0.10\columnwidth}\centering
1
\end{minipage} & \begin{minipage}[t]{0.39\columnwidth}\raggedright
candidate, announcement
\end{minipage} & \begin{minipage}[t]{0.15\columnwidth}\raggedright
\textbf{episodic}
\end{minipage}
\\\addlinespace
\begin{minipage}[t]{0.10\columnwidth}\centering
2
\end{minipage} & \begin{minipage}[t]{0.39\columnwidth}\raggedright
health insurance
\end{minipage} & \begin{minipage}[t]{0.15\columnwidth}\raggedright
\textbf{thematic}
\end{minipage}
\\\addlinespace
\begin{minipage}[t]{0.10\columnwidth}\centering
3
\end{minipage} & \begin{minipage}[t]{0.39\columnwidth}\raggedright
republican party politics
\end{minipage} & \begin{minipage}[t]{0.15\columnwidth}\raggedright
\textbf{episodic}
\end{minipage}
\\\addlinespace
\begin{minipage}[t]{0.10\columnwidth}\centering
4
\end{minipage} & \begin{minipage}[t]{0.39\columnwidth}\raggedright
\emph{(unparsable)}
\end{minipage} & \begin{minipage}[t]{0.15\columnwidth}\raggedright
\emph{NA}
\end{minipage}
\\\addlinespace
\begin{minipage}[t]{0.10\columnwidth}\centering
5
\end{minipage} & \begin{minipage}[t]{0.39\columnwidth}\raggedright
democratic call to action
\end{minipage} & \begin{minipage}[t]{0.15\columnwidth}\raggedright
\textbf{episodic}
\end{minipage}
\\\addlinespace
\begin{minipage}[t]{0.10\columnwidth}\centering
6
\end{minipage} & \begin{minipage}[t]{0.39\columnwidth}\raggedright
democratic call for donations
\end{minipage} & \begin{minipage}[t]{0.15\columnwidth}\raggedright
\textbf{episodic}
\end{minipage}
\\\addlinespace
\begin{minipage}[t]{0.10\columnwidth}\centering
7
\end{minipage} & \begin{minipage}[t]{0.39\columnwidth}\raggedright
\emph{(unparsable)}
\end{minipage} & \begin{minipage}[t]{0.15\columnwidth}\raggedright
\emph{NA}
\end{minipage}
\\\addlinespace
\begin{minipage}[t]{0.10\columnwidth}\centering
8
\end{minipage} & \begin{minipage}[t]{0.39\columnwidth}\raggedright
church and state separation
\end{minipage} & \begin{minipage}[t]{0.15\columnwidth}\raggedright
\textbf{thematic}
\end{minipage}
\\\addlinespace
\begin{minipage}[t]{0.10\columnwidth}\centering
9
\end{minipage} & \begin{minipage}[t]{0.39\columnwidth}\raggedright
democratic party politics
\end{minipage} & \begin{minipage}[t]{0.15\columnwidth}\raggedright
\textbf{episodic}
\end{minipage}
\\\addlinespace
\begin{minipage}[t]{0.10\columnwidth}\centering
10
\end{minipage} & \begin{minipage}[t]{0.39\columnwidth}\raggedright
\emph{(unparseable)}
\end{minipage} & \begin{minipage}[t]{0.15\columnwidth}\raggedright
\emph{NA}
\end{minipage}
\\\addlinespace
\bottomrule
\addlinespace
\caption{\label{tab:dt}Topics on websites tweeted by Democrats.}
\end{longtable}

\begin{longtable}[c]{@{}cll@{}}
\toprule\addlinespace
\begin{minipage}[b]{0.11\columnwidth}\centering
Number
\end{minipage} & \begin{minipage}[b]{0.38\columnwidth}\raggedright
Content
\end{minipage} & \begin{minipage}[b]{0.15\columnwidth}\raggedright
Aspect
\end{minipage}
\\\addlinespace
\midrule\endhead
\begin{minipage}[t]{0.11\columnwidth}\centering
1
\end{minipage} & \begin{minipage}[t]{0.38\columnwidth}\raggedright
republican call to action
\end{minipage} & \begin{minipage}[t]{0.15\columnwidth}\raggedright
\textbf{episodic}
\end{minipage}
\\\addlinespace
\begin{minipage}[t]{0.11\columnwidth}\centering
2
\end{minipage} & \begin{minipage}[t]{0.38\columnwidth}\raggedright
national unemployment
\end{minipage} & \begin{minipage}[t]{0.15\columnwidth}\raggedright
\textbf{thematic}
\end{minipage}
\\\addlinespace
\begin{minipage}[t]{0.11\columnwidth}\centering
3
\end{minipage} & \begin{minipage}[t]{0.38\columnwidth}\raggedright
candidate, announcement
\end{minipage} & \begin{minipage}[t]{0.15\columnwidth}\raggedright
\textbf{episodic}
\end{minipage}
\\\addlinespace
\begin{minipage}[t]{0.11\columnwidth}\centering
4
\end{minipage} & \begin{minipage}[t]{0.38\columnwidth}\raggedright
obama care is unpopular
\end{minipage} & \begin{minipage}[t]{0.15\columnwidth}\raggedright
\textbf{thematic}
\end{minipage}
\\\addlinespace
\begin{minipage}[t]{0.11\columnwidth}\centering
5
\end{minipage} & \begin{minipage}[t]{0.38\columnwidth}\raggedright
republican call to action
\end{minipage} & \begin{minipage}[t]{0.15\columnwidth}\raggedright
\textbf{episodic}
\end{minipage}
\\\addlinespace
\begin{minipage}[t]{0.11\columnwidth}\centering
6
\end{minipage} & \begin{minipage}[t]{0.38\columnwidth}\raggedright
obama care is expensive
\end{minipage} & \begin{minipage}[t]{0.15\columnwidth}\raggedright
\textbf{thematic}
\end{minipage}
\\\addlinespace
\begin{minipage}[t]{0.11\columnwidth}\centering
7
\end{minipage} & \begin{minipage}[t]{0.38\columnwidth}\raggedright
republican call for donations
\end{minipage} & \begin{minipage}[t]{0.15\columnwidth}\raggedright
\textbf{episodic}
\end{minipage}
\\\addlinespace
\begin{minipage}[t]{0.11\columnwidth}\centering
8
\end{minipage} & \begin{minipage}[t]{0.38\columnwidth}\raggedright
\emph{(unparsable)}
\end{minipage} & \begin{minipage}[t]{0.15\columnwidth}\raggedright
\emph{NA}
\end{minipage}
\\\addlinespace
\begin{minipage}[t]{0.11\columnwidth}\centering
9
\end{minipage} & \begin{minipage}[t]{0.38\columnwidth}\raggedright
arguments against obamacare
\end{minipage} & \begin{minipage}[t]{0.15\columnwidth}\raggedright
\textbf{thematic}
\end{minipage}
\\\addlinespace
\begin{minipage}[t]{0.11\columnwidth}\centering
10
\end{minipage} & \begin{minipage}[t]{0.38\columnwidth}\raggedright
republican call for support
\end{minipage} & \begin{minipage}[t]{0.15\columnwidth}\raggedright
\textbf{episodic}
\end{minipage}
\\\addlinespace
\bottomrule
\addlinespace
\caption{\label{tab:rt}Topics on websites tweeted by Republicans.}
\end{longtable}

As LDA provides posterior probabilities of belonging to each topic for
each website ($M$), as depicted in Figure \ref{fig:topic-heatmap},
thematicity and episodicity scores were computed for each tweet linking
to a website. This episodicity (thematicity) score was obtained by
summing up those elements $m_{ik}$ of $M$ where $k$ is an element of the
set of all episodic (thematic) topics $K$:

\begin{center}

$e_i = \sum_{k \in K} m_{ik}$

\end{center}

\begin{figure}[htbp]
\centering
\includegraphics{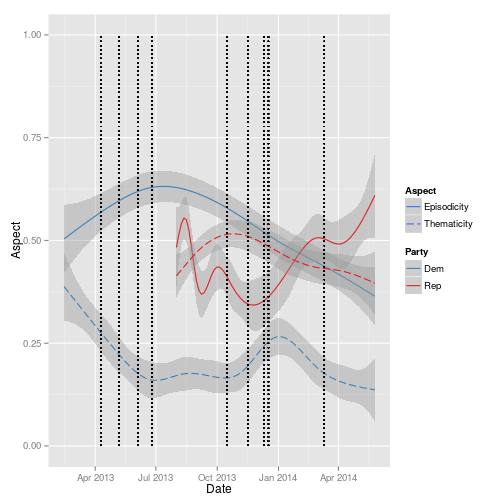}
\caption{\label{fig:et-overtime}The development of the prevalence of
episodic and thematic content over time for both parties. Trend lines
are GAM-based and elections to House or Senate are indicated by dotted
verticals.}
\end{figure}

Therefore, episodicity (thematicity) scores range from 0 to 1 with 0
being a website without any episodic (thematic) topics and 1 a website
containing solely episodic (thematic) topics. The development of
episodicity and thematicity over time for both parties is given in
Figure \ref{fig:et-overtime}. It becomes evident that party strategies
with respect to the prevalence of are not only divergent but also
changing over time.

Starting from these descriptive observations, I will now present a model
linking episodicity and thematicity with the reshare count.

\section{Results}\label{results}

In order to establish any association between episodicity, thematicity
and the reshare count of a message, a generalized linear model is
considered. Based the findings in Hochreiter and Waldhauser (2013), the
model contains control variables for time of day and message length. An
additional control variable relating to the distance from a tweet's
posting to the closest election is also included in the model.\footnote{In
  order to stabilize the variance in that distance, the square root of
  the term was considered.} As count-based models require constant
observational frames, an offset term computed from the number of
followers for the respective party account ($n_p$) times the age of the
message ($a$) was included as well:

\begin{center}
$o = n_p \times a$
\end{center}

The initial model contained the first order terms for party membership,
episodicity, thematicity, time of day, message length and the election
proximity. All these terms were complemented with interactions with the
party variable. Finally, thematicity and episodicity interactions with
party were included as well.

Starting with this complete model, insignificant interactions were
removed, one at a time. This led to the removal of the party---message
length and party---thematicity interactions. Table \ref{tab:rms}
contains the test results from the model selection step.

\begin{longtable}[c]{@{}ccccc@{}}
\toprule\addlinespace
\begin{minipage}[b]{0.18\columnwidth}\centering
2 x log-lik.
\end{minipage} & \begin{minipage}[b]{0.08\columnwidth}\centering
Test
\end{minipage} & \begin{minipage}[b]{0.06\columnwidth}\centering
df
\end{minipage} & \begin{minipage}[b]{0.13\columnwidth}\centering
LR stat.
\end{minipage} & \begin{minipage}[b]{0.18\columnwidth}\centering
$\Pr(\chi^2)$
\end{minipage}
\\\addlinespace
\midrule\endhead
\begin{minipage}[t]{0.18\columnwidth}\centering
-33964
\end{minipage} & \begin{minipage}[t]{0.08\columnwidth}\centering
1 vs 2
\end{minipage} & \begin{minipage}[t]{0.06\columnwidth}\centering
1
\end{minipage} & \begin{minipage}[t]{0.13\columnwidth}\centering
1.95
\end{minipage} & \begin{minipage}[t]{0.18\columnwidth}\centering
0.1626
\end{minipage}
\\\addlinespace
\begin{minipage}[t]{0.18\columnwidth}\centering
-33963
\end{minipage} & \begin{minipage}[t]{0.08\columnwidth}\centering
2 vs 3
\end{minipage} & \begin{minipage}[t]{0.06\columnwidth}\centering
1
\end{minipage} & \begin{minipage}[t]{0.13\columnwidth}\centering
0.8676
\end{minipage} & \begin{minipage}[t]{0.18\columnwidth}\centering
0.3516
\end{minipage}
\\\addlinespace
\bottomrule
\addlinespace
\caption{\label{tab:rms}Likelihood ratio tests confirming dropping of
party---message length and party---thematicity interactions from the
model.}
\end{longtable}

Owing to the count nature of the dependent variable, Poisson regression
or a negative-binomial model must be used. Based on a likelihood-ratio
test comparing the Poisson and negative binomial versions of the model,
$\chi^2(14, N = 3592) = 138,045.92, p = 0$, equality of conditional mean
and variance cannot be assumed and a negative binomial model is
therefore more appropriate.

The final model's estimates, standard errors and p-values are given in
Table \ref{tab:m1.2}.

\begin{longtable}[c]{@{}cccc@{}}
\toprule\addlinespace
\begin{minipage}[b]{0.30\columnwidth}\centering
~
\end{minipage} & \begin{minipage}[b]{0.15\columnwidth}\centering
$\beta_i$
\end{minipage} & \begin{minipage}[b]{0.07\columnwidth}\centering
SE
\end{minipage} & \begin{minipage}[b]{0.11\columnwidth}\centering
p-value
\end{minipage}
\\\addlinespace
\midrule\endhead
\begin{minipage}[t]{0.30\columnwidth}\centering
\textbf{(Intercept)}
\end{minipage} & \begin{minipage}[t]{0.15\columnwidth}\centering
-14.45
\end{minipage} & \begin{minipage}[t]{0.07\columnwidth}\centering
0.139
\end{minipage} & \begin{minipage}[t]{0.11\columnwidth}\centering
0
\end{minipage}
\\\addlinespace
\begin{minipage}[t]{0.30\columnwidth}\centering
\textbf{Party (1=R)}
\end{minipage} & \begin{minipage}[t]{0.15\columnwidth}\centering
-0.142
\end{minipage} & \begin{minipage}[t]{0.07\columnwidth}\centering
0.095
\end{minipage} & \begin{minipage}[t]{0.11\columnwidth}\centering
0.136
\end{minipage}
\\\addlinespace
\begin{minipage}[t]{0.30\columnwidth}\centering
\textbf{Episodicity}
\end{minipage} & \begin{minipage}[t]{0.15\columnwidth}\centering
-1.132
\end{minipage} & \begin{minipage}[t]{0.07\columnwidth}\centering
0.125
\end{minipage} & \begin{minipage}[t]{0.11\columnwidth}\centering
0
\end{minipage}
\\\addlinespace
\begin{minipage}[t]{0.30\columnwidth}\centering
\textbf{Thematicity}
\end{minipage} & \begin{minipage}[t]{0.15\columnwidth}\centering
-0.73
\end{minipage} & \begin{minipage}[t]{0.07\columnwidth}\centering
0.134
\end{minipage} & \begin{minipage}[t]{0.11\columnwidth}\centering
0
\end{minipage}
\\\addlinespace
\begin{minipage}[t]{0.30\columnwidth}\centering
\textbf{Retweet (1=Yes)}
\end{minipage} & \begin{minipage}[t]{0.15\columnwidth}\centering
1.526
\end{minipage} & \begin{minipage}[t]{0.07\columnwidth}\centering
0.164
\end{minipage} & \begin{minipage}[t]{0.11\columnwidth}\centering
0
\end{minipage}
\\\addlinespace
\begin{minipage}[t]{0.30\columnwidth}\centering
\textbf{Time of Day}
\end{minipage} & \begin{minipage}[t]{0.15\columnwidth}\centering
0.039
\end{minipage} & \begin{minipage}[t]{0.07\columnwidth}\centering
0.009
\end{minipage} & \begin{minipage}[t]{0.11\columnwidth}\centering
0
\end{minipage}
\\\addlinespace
\begin{minipage}[t]{0.30\columnwidth}\centering
\textbf{Message Length}
\end{minipage} & \begin{minipage}[t]{0.15\columnwidth}\centering
0.003
\end{minipage} & \begin{minipage}[t]{0.07\columnwidth}\centering
0.001
\end{minipage} & \begin{minipage}[t]{0.11\columnwidth}\centering
0
\end{minipage}
\\\addlinespace
\begin{minipage}[t]{0.30\columnwidth}\centering
\textbf{$\sqrt{Proxi.}$}
\end{minipage} & \begin{minipage}[t]{0.15\columnwidth}\centering
0.229
\end{minipage} & \begin{minipage}[t]{0.07\columnwidth}\centering
0.02
\end{minipage} & \begin{minipage}[t]{0.11\columnwidth}\centering
0
\end{minipage}
\\\addlinespace
\begin{minipage}[t]{0.30\columnwidth}\centering
\textbf{Episodicity (Rep)}
\end{minipage} & \begin{minipage}[t]{0.15\columnwidth}\centering
0.241
\end{minipage} & \begin{minipage}[t]{0.07\columnwidth}\centering
0.089
\end{minipage} & \begin{minipage}[t]{0.11\columnwidth}\centering
0.007
\end{minipage}
\\\addlinespace
\begin{minipage}[t]{0.30\columnwidth}\centering
\textbf{Retweet (Rep)}
\end{minipage} & \begin{minipage}[t]{0.15\columnwidth}\centering
-1.104
\end{minipage} & \begin{minipage}[t]{0.07\columnwidth}\centering
0.218
\end{minipage} & \begin{minipage}[t]{0.11\columnwidth}\centering
0
\end{minipage}
\\\addlinespace
\begin{minipage}[t]{0.30\columnwidth}\centering
\textbf{Time of Day (Rep)}
\end{minipage} & \begin{minipage}[t]{0.15\columnwidth}\centering
-0.029
\end{minipage} & \begin{minipage}[t]{0.07\columnwidth}\centering
0.012
\end{minipage} & \begin{minipage}[t]{0.11\columnwidth}\centering
0.017
\end{minipage}
\\\addlinespace
\begin{minipage}[t]{0.30\columnwidth}\centering
\textbf{Episodicity $\times \sqrt{Proxi.}$}
\end{minipage} & \begin{minipage}[t]{0.15\columnwidth}\centering
-0.079
\end{minipage} & \begin{minipage}[t]{0.07\columnwidth}\centering
0.024
\end{minipage} & \begin{minipage}[t]{0.11\columnwidth}\centering
0.001
\end{minipage}
\\\addlinespace
\begin{minipage}[t]{0.30\columnwidth}\centering
\textbf{Thematicity $\times \sqrt{Proxi.}$}
\end{minipage} & \begin{minipage}[t]{0.15\columnwidth}\centering
-0.11
\end{minipage} & \begin{minipage}[t]{0.07\columnwidth}\centering
0.026
\end{minipage} & \begin{minipage}[t]{0.11\columnwidth}\centering
0
\end{minipage}
\\\addlinespace
\bottomrule
\addlinespace
\caption{\label{tab:m1.2}Final model coefficients.}
\end{longtable}

\section{Discussion}\label{discussion}

Akin to the more familiar odds ratios from logistic regression, count
data sports incidence rate ratios as interpretation of its exponentiated
coefficients. These IRRs can be found in Table \ref{tab:irr}.

\begin{longtable}[c]{@{}cc@{}}
\toprule\addlinespace
\begin{minipage}[b]{0.31\columnwidth}\centering
~
\end{minipage} & \begin{minipage}[b]{0.21\columnwidth}\centering
$\exp \beta_i$
\end{minipage}
\\\addlinespace
\midrule\endhead
\begin{minipage}[t]{0.31\columnwidth}\centering
\textbf{(Intercept)}
\end{minipage} & \begin{minipage}[t]{0.21\columnwidth}\centering
0
\end{minipage}
\\\addlinespace
\begin{minipage}[t]{0.31\columnwidth}\centering
\textbf{Party (1=R)}
\end{minipage} & \begin{minipage}[t]{0.21\columnwidth}\centering
0.867
\end{minipage}
\\\addlinespace
\begin{minipage}[t]{0.31\columnwidth}\centering
\textbf{Episodicity}
\end{minipage} & \begin{minipage}[t]{0.21\columnwidth}\centering
0.322
\end{minipage}
\\\addlinespace
\begin{minipage}[t]{0.31\columnwidth}\centering
\textbf{Thematicity}
\end{minipage} & \begin{minipage}[t]{0.21\columnwidth}\centering
0.482
\end{minipage}
\\\addlinespace
\begin{minipage}[t]{0.31\columnwidth}\centering
\textbf{Retweet (1=Yes)}
\end{minipage} & \begin{minipage}[t]{0.21\columnwidth}\centering
4.599
\end{minipage}
\\\addlinespace
\begin{minipage}[t]{0.31\columnwidth}\centering
\textbf{Time of Day}
\end{minipage} & \begin{minipage}[t]{0.21\columnwidth}\centering
1.04
\end{minipage}
\\\addlinespace
\begin{minipage}[t]{0.31\columnwidth}\centering
\textbf{Message Length}
\end{minipage} & \begin{minipage}[t]{0.21\columnwidth}\centering
1.003
\end{minipage}
\\\addlinespace
\begin{minipage}[t]{0.31\columnwidth}\centering
\textbf{$\sqrt{Proxi.}$}
\end{minipage} & \begin{minipage}[t]{0.21\columnwidth}\centering
1.257
\end{minipage}
\\\addlinespace
\begin{minipage}[t]{0.31\columnwidth}\centering
\textbf{Episodicity (Rep)}
\end{minipage} & \begin{minipage}[t]{0.21\columnwidth}\centering
1.272
\end{minipage}
\\\addlinespace
\begin{minipage}[t]{0.31\columnwidth}\centering
\textbf{Retweet (Rep)}
\end{minipage} & \begin{minipage}[t]{0.21\columnwidth}\centering
0.332
\end{minipage}
\\\addlinespace
\begin{minipage}[t]{0.31\columnwidth}\centering
\textbf{Time of Day (Rep)}
\end{minipage} & \begin{minipage}[t]{0.21\columnwidth}\centering
0.972
\end{minipage}
\\\addlinespace
\begin{minipage}[t]{0.31\columnwidth}\centering
\textbf{Episodicity $\times \sqrt{Proxi.}$}
\end{minipage} & \begin{minipage}[t]{0.21\columnwidth}\centering
0.924
\end{minipage}
\\\addlinespace
\begin{minipage}[t]{0.31\columnwidth}\centering
\textbf{Thematicity $\times \sqrt{Proxi.}$}
\end{minipage} & \begin{minipage}[t]{0.21\columnwidth}\centering
0.896
\end{minipage}
\\\addlinespace
\bottomrule
\addlinespace
\caption{\label{tab:irr}Incidence rate ratios for the model coefficients
reflecting the expected change in retweet rates for one unit increases.}
\end{longtable}

From these results, I am going to discuss the more remarkable ones here.

\begin{enumerate}
\def\labelenumi{\arabic{enumi}.}
\item
  There is a quite strong difference between reshares Republican and
  Democrats can expect, even when controlling for different user base
  sizes: for a tweet that would have generated 100 reshares if it had
  originated with the Democrats, the Republicans can only expect 87.

  Obviously, Democrats lead social media usage not only by the number of
  followers but also by their (the followers') dedication to the medium.
\item
  Episodic and thematic content enjoy virtually the same popularity. It
  is also evident, that the less thematic \emph{or} episodic a message
  is, the higher the expected reshare count. While surprising at first,
  this is an artifact introduced by unparsable content on the websites.
  It is entirely conceivable that images -- for their simplicity and
  easy to understand punchlines -- are more popular all together than
  textual content.

  When comparing textual content directly, it becomes evident that
  thematic content is almost 150 percent as popular as episodic content.
  Interestingly, this difference is only to be found in Democratic
  tweets and all but disappears for tweets sent from the Republican
  account. These differences are contrary to the predictions of Iyengar,
  Norpoth, and Hahn (2004). There, consumers would \emph{want} episodic,
  horserace news. In this data set, however, users in general don't like
  textual content. Among textual content they do prefer thematic over
  episodic news.
\item
  Finally, there are interesting effects dependent on the proximity of a
  message to election day. In general, the expected reshare count
  increases by 25 percent for every additional square root day away from
  an election. This effect is toned done somewhat for episodic or
  thematic content.
\end{enumerate}

\section{Conclusion}\label{conclusion}

In this paper I have been looking at the way both US political parties
make use of social media. The example case for this study was Twitter
and I analyzed the last 3,200 tweets that originated from the main party
accounts along with any websites they might point to. I've concentrated
on tracing how thematic and episodic content is being distributed
(differently) via Twitter. To that end, I've text-mined any websites
that were mentioned in those tweets and used LDA topic models to extract
the topics that occur in these websites. Using Iyengar's definition of
episodic or horserace news, I classified all (parsable) topics to be
either episodic or thematic. Finally, generalized linear models are used
to model the relationship between the retweets a message receives and
episodicity, thematicity and some controlling covariates.

The results in parts contradict what theory would have predicted, in
that thematic news is more popular than episodic one. Iyengar, Norpoth,
and Hahn (2004) argue that mass media focuses on episodic news, because
that is what consumers demand. I, therefore, tentatively conclude that
social media would have a potential as a public sphere, supporting
arguments in political deliberation.

While these results are only preliminary and from a very limited data
set, they are informative in the sense that they lead us to question the
long-held assumption that market forces lead to episodic news coverage.
Clearly, more in-depth and broader analysis is needed to grow confidence
in this challenge.

\section{References}\label{references}

Blumler, Jay G, and Dennis Kavanagh. 1999. ``The Third Age of Political
Communication: Influences and Features.'' \emph{Political Communication}
16 (3): 209--230.

Cohen, Joshua. 2003. ``Deliberation and Democratic Legitimacy.'' In
\emph{Debates in Contemporary Political Philosophy. an Anthology},
edited by Derek Matravers and Jon Pike, 342--360. New York: Routledge.

Dahlberg, Lincoln. 2001. ``The Internet and Democratic Discourse:
Exploring the Prospects of Online Deliberative Forums Extending the
Public Sphere.'' \emph{Information, Communication \& Society} 4 (4):
615--633.

Dahlgren, Peter. 2005. ``The Internet, Public Spheres, and Political
Communication: Dispersion and Deliberation.'' \emph{Political
Communication} 22 (2): 147--162.

Elster, Jon, ed. 1998. \emph{Deliberative Democracy}. Cambridge; New
York; Melbourne: Cambridge University Press.

Habermas, Jürgen. 1991. \emph{The Structural Transformation of the
Public Sphere: An Inquiry into a Category of Bourgeois Society}. MIT
Press.

---------. 1994. ``Three Normative Models of Democracy.''
\emph{Constellations: An International Journal of Critical and
Democratic Theory} 1 (1): 1--10.

---------. 2006. ``Political Communication in Media Society: Does
Democracy Still Enjoy an Epistemic Dimension? The Impact of Normative
Theory on Empirical Research.'' \emph{Communication Theory} 16 (4):
411--426.

Hjarvard, Stig. 2008. ``The Mediatization of Society: A Theory of the
Media as Agents of Social and Cultural Change.'' \emph{Nordicom Review:
Nordic Research on Media \& Communication} 29 (2): 105--134.

Hochreiter, Ronald, and Christoph Waldhauser. 2013. ``A Stochastic
Simulation of the Decision to Retweet.'' In \emph{Algorithmic Decision
Theory}, 221--229. Springer.

Iyengar, Shanto, and Donald R Kinder. 2010. \emph{News That Matters:
Television and American Opinion}. Chicago, Illinois: University of
Chicago Press.

Iyengar, Shanto, and Adam F Simon. 2000. ``New Perspectives and Evidence
on Political Communication and Campaign Effects.'' \emph{Annual Review
of Psychology} 51 (1): 149--169.

Iyengar, Shanto, Robert C Luskin, and James Fishkin. 2004.
``Deliberative Public Opinion in Presidential Primaries: Evidence from
the Online Deliberative Poll.''

Iyengar, Shanto, Helmut Norpoth, and Kyu S Hahn. 2004. ``Consumer Demand
for Election News: The Horserace Sells.'' \emph{Journal of Politics} 66
(1): 157--175.

Koopmans, Ruud, and Jessica Erbe. 2004. ``Towards a European Public
Sphere? Vertical and Horizontal Dimensions of Europeanized Political
Communication.'' \emph{Innovation: The European Journal of Social
Science Research} 17 (2): 97--118.

Meyer, Thomas, and Lewis P Hinchman. 2002. \emph{Media Democracy: How
the Media Colonize Politics}. Cambridge: Polity Press.

Pettersson, David, and Petter Karlström. 2011. ``Reputation as a
Product: Politicians in Social Media.''

Raymond, Eric. 1999. ``The Cathedral and the Bazaar.'' \emph{Knowledge,
Technology \& Policy} 12 (3): 23--49.

Rheingold, Howard. 1993. \emph{The Virtual Community: Homesteading on
the Electronic Frontier}. Reading, Massachusetts: Addison-Wesley
Longman.

Risse, Thomas, and Marianne Van de Steeg. 2003. ``An Emerging European
Public Sphere? Empirical Evidence and Theoretical Clarifications.''

Schulz, Winfried. 2004. ``Reconstructing Mediatization as an Analytical
Concept.'' \emph{European Journal of Communication} 19 (1): 87--101.

Strömbäck, Jesper. 2008. ``Four Phases of Mediatization: An Analysis of
the Mediatization of Politics.'' \emph{The International Journal of
Press/Politics} 13 (3): 228--246.

Strömbäck, Jesper, and Frank Esser. 2009. ``Shaping Politics:
Mediatization and Media Interventionism: Concept, Changes,
Consequences.'' In \emph{Mediatization}, edited by Knut Lundby,
205--224. New York: Peter Lang.

Traugott, Michael W, and Paul J Lavrakas. 2008. \emph{The Voter's Guide
to Election Polls}. New York: Rowman \& Littlefield.

\end{document}